\documentclass[aps,prl,reprint,superscriptaddress]{revtex4-2}

\usepackage{float}
\usepackage{stmaryrd}
\usepackage{amsmath} 

\usepackage{gensymb}

\usepackage{textcomp}
\usepackage{array}

\usepackage{graphicx}
\usepackage{amsmath}
\usepackage{amssymb}
\usepackage{epstopdf}
\usepackage{bm}

\usepackage[colorinlistoftodos]{todonotes}
\usepackage{siunitx}
\usepackage[absolute,overlay]{textpos}

\usepackage{upgreek}
\usepackage[colorlinks=true, citecolor={black}, urlcolor={blue!60!black}, linkcolor = {black}]{hyperref}
\usepackage[nameinlink]{cleveref}
\Crefname{figure}{Fig.}{Figs.}

\usepackage{comment}
\usepackage{bold-extra}

\usepackage[normalem]{ulem}

\makeatletter
\newcommand{\fmarki}{*}
\newcommand{\fmarkii}{\ensuremath{\dagger}}
\newcommand{\fmarkiii}{\ensuremath{\ddagger}}
\newcommand{\fmarkiv}{\ensuremath{\mathsection}}
\newcommand{\fmarkv}{\ensuremath{\mathparagraph}}
\newcommand{\fmarkvi}{\ensuremath{\|}}
\newcommand{\fmarkvii}{**}
\newcommand{\fmarkviii}{\ensuremath{\dagger\dagger}}
\newcommand{\fmarkix}{\ensuremath{\ddagger\ddagger}}
                
\def\@fnsymbol#1{{\ifcase#1\or \fmarki\or \fmarkii\or \fmarkiii\or \fmarkiv\or \fmarkv\or \fmarkvi\or \fmarkvii\or \fmarkviii\or \fmarkix \else\@ctrerr\fi}}
\makeatother

\renewcommand{\fmarki}{\ensuremath{\dagger}}
\renewcommand{\fmarkii}{*}

\renewcommand{\abstractname}{ABSTRACT}

\begin{document}

\title{\textcolor{black}{Triplet correlations in Cooper pair splitters realized in a two-dimensional electron gas}}










\author{Qingzhen Wang}
\affiliation{QuTech and Kavli Institute of Nanoscience, Delft University of Technology, Delft, 2600 GA, The Netherlands}

\author{Sebastiaan L.D. ten Haaf}
\affiliation{QuTech and Kavli Institute of Nanoscience, Delft University of Technology, Delft, 2600 GA, The Netherlands}

\author{Ivan Kulesh}
\affiliation{QuTech and Kavli Institute of Nanoscience, Delft University of Technology, Delft, 2600 GA, The Netherlands}

\author{Di~Xiao}
\affiliation{Department of Physics and Astronomy, Purdue University, West Lafayette, 47907, Indiana, USA}

\author{Candice Thomas}
\affiliation{Department of Physics and Astronomy, Purdue University, West Lafayette, 47907, Indiana, USA}

\author{Michael J. Manfra}
\affiliation{Department of Physics and Astronomy, Purdue University, West Lafayette, 47907, Indiana, USA}
\affiliation{Elmore School of Electrical and Computer Engineering, ~Purdue University, West Lafayette, 47907, Indiana, USA}
\affiliation{School of Materials Engineering, Purdue University, West Lafayette, 47907, Indiana, USA}
\affiliation{Microsoft Quantum Lab, West Lafayette, 47907, Indiana, USA}

\author{Srijit Goswami*}
\affiliation{QuTech and Kavli Institute of Nanoscience, Delft University of Technology, Delft, 2600 GA, The Netherlands}

\begin{abstract}
\begin{center}
$\dagger$These authors contributed equally to this work.

* \url{s.goswami@tudelft.nl}

    \textbf{\abstractname}
\end{center}
Cooper pairs occupy the ground state of superconductors and are \textcolor{black}{typically} composed of maximally entangled electrons with opposite spin. 
In order to study the spin and entanglement properties of these  electrons, one must separate them spatially via a process known as Cooper pair splitting (CPS). 
Here we provide the first demonstration of CPS in a semiconductor two-dimensional electron gas (2DEG). 
By coupling two quantum dots to a superconductor-semiconductor hybrid region we achieve efficient Cooper pair splitting, and clearly distinguish it from other local and non-local processes. 
When the spin degeneracy of the dots is lifted, they can be operated as spin-filters to obtain information about the spin of the electrons forming the Cooper pair. 
Not only do we observe a near perfect splitting of Cooper pairs into opposite-spin electrons (i.e. conventional singlet pairing), but also into equal-spin electrons, thus achieving triplet \textcolor{black}{correlations} between the quantum dots. 
Importantly, the exceptionally large spin-orbit interaction in our 2DEGs results in a strong triplet \textcolor{black}{component}, comparable in amplitude to the singlet pairing. 
The demonstration of CPS in a scalable and flexible platform provides a credible route to study on-chip entanglement and topological superconductivity in the form of artificial Kitaev chains.
\end{abstract}
\maketitle

\section{Introduction}
Coupling two normal leads to a superconductor can give rise to non-local transport processes directly involving both leads. 
Two opposite-spin electrons from a Cooper pair in the superconductor can be split into the leads via a process known as Cooper pair splitting (CPS). 
The dominant transport mechanism that gives rise to CPS is crossed Andreev reflection (CAR), whereby \textcolor{black}{a higher order process} allows two electrons to be injected simultaneously into the superconductor to form a Cooper pair.
Additionally, a single electron can tunnel through the superconductor from one lead to the other through a process known as elastic co-tunnelling (ECT). 
The ability to control these processes has important implications for two distinct fields.
Firstly, efficient CPS can be used to generate spatially separated entangled electrons, that can be used to perform a Bell test~\cite{Recher2001,Chtchelkatchev2002,Braunecker2013,Klobus2014,Brange_PRL_2017,Busz2017}. 
Secondly, in the context of topological superconductivity, it has been shown that CAR and ECT are crucial ingredients required to implement a Kitaev chain~\cite{Kitaev_2001} using quantum dot-superconductor hybrids~\cite{Sau2012,Flensberg2012}.
\begin{figure*}[t!]
\centering
\includegraphics[width=1\textwidth]{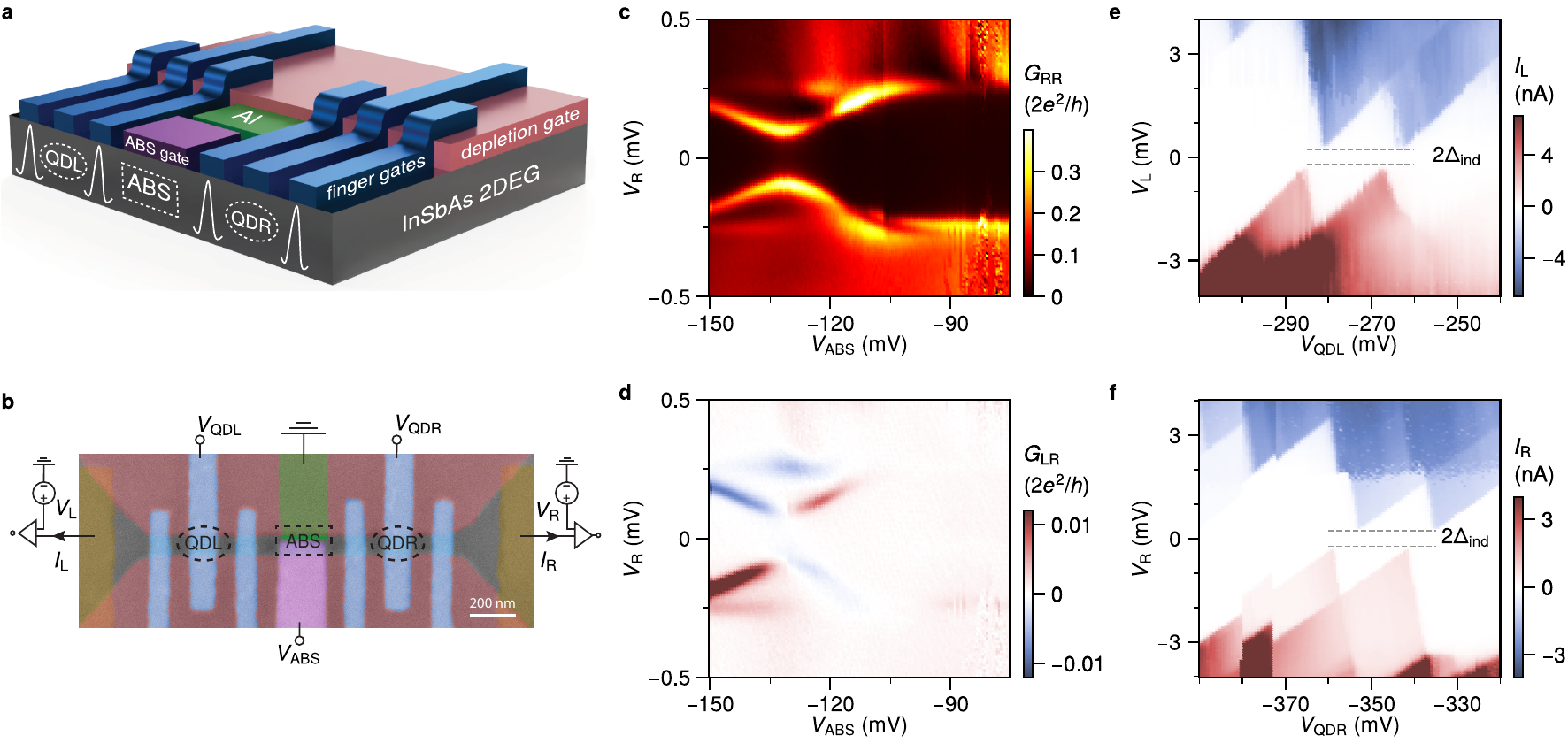}
\caption{\textbf{Basic device characterization.} \textbf{(a)} A 3D illustration of the device. The two quantum dots (QDL and QDR), and the region hosting ABSs are indicated. For clarity, the gate-dielectric layers are not shown. \textbf{(b)} False-color scanning electron micrograph of Device~1, including a schematic of the circuit diagram for three-terminal measurements. Tunneling spectroscopy measurements showing \textbf{(c)} local conductance $G_\mathrm{RR}$ and \textbf{(d)} non-local conductance $G_\mathrm{LR}$ as a function of the ABS gate voltage $V_\mathrm{ABS}$ and right bias voltage $V_\mathrm{R}$. Coulomb diamonds of the QDs are measured for \textbf{(e)} QDL and \textbf{(f)} QDR.}
\label{fig1}
\end{figure*}

CPS has been studied in various mesoscopic systems coupled to superconductors, such as semiconductor nanowires~\cite{Hofstetter2009,Hofstetter2011,Das2012,Deacon2015}, carbon nanotubes~\cite{Herrmann2010,Schindele_PRL_2012}, and graphene~\cite{Tan2015}. Quantum dots (QDs) are generally added between the leads and the superconductor.
The charging energy of the QDs ensures that electrons forming a Cooper pair preferentially split into separate dots, rather than  occupying levels in the same dot. 
This results in correlated electrical currents at the two normal leads.
It has thus far been challenging to independently measure the relevant virtual processes (i.e. ECT and CAR) and isolate them from local processes, such as normal Andreev reflection or direct tunnelling via sub-gap states. 
In a set of recent studies on hybrid nanowires, it was shown that these challenges could be overcome to create a highly efficient Cooper pair splitter~\cite{Wang2022} and to realize a minimal Kitaev chain~\cite{Dvir2022}. 
A key idea is that the QDs were coupled via extended Andreev bound states (ABSs) in the semiconductor-superconductor hybrid~\cite{ChunXiao2022,Tsintzis2022,bordin2022controlled}, rather than the continuum above the superconducting gap.
Therefore, by controlling the ABS energy with electrostatic gates, it was possible to tune the relative amplitudes of ECT and CAR. 
These developments pave the way for more advanced  experiments, where the geometrical constraints of 1D systems will pose restrictions on the complexity of possible devices. 
An ideal platform to overcome these restrictions are semiconductor 2DEGs. Not only do they offer flexibility in device design, but also serve as a scalable platform to create and manipulate topologically protected Majorana bound states in  artificial Kitaev chains.

We demonstrate here for the first time the observation of Cooper pair splitting in a 2D semiconductor platform. 
This is achieved by coupling two quantum dots via a hybrid proximitized section in an InSbAs 2DEG. 
By applying an external magnetic field, we polarize the spins of the QDs, allowing us to use them as spin-filters. 
This, in combination with highly efficient CPS, allows us to accurately resolve the spin of the electrons involved in CAR and ECT. 
\textcolor{black}{The large spin-orbit coupling in our 2DEGs, in combination with the device dimensions, results in significant spin precession for the electrons.
Importantly, we show that this leads to strong equal-spin CAR currents that are of similar amplitude to the conventional opposite-spin processes.}
Through rotation of the magnetic field angle relative to the spin-orbit field, we show that the ECT and CAR processes can be tuned to equal amplitudes, satisfying a key requirement for realizing a Kitaev chain in semiconductor-superconductor hybrids.

\section{Results}
\subsection{Device and characterization}
Devices are fabricated on an InSbAs 2DEG with epitaxial aluminum grown by molecular beam epitaxy. 
This material has been established to have a low effective mass, high g-factor and large spin-orbit coupling \cite{Moehle2021, Metti_PRB_2022}. 
\Cref{fig1}a,b illustrate the device structure together with the three-terminal measurement circuit.
The two depletion gates (pink) define a quasi-1D channel of about \SI{150}{nm}, contacted on each side with normal leads.
The middle of the channel is proximitized via a \SI{150}{nm}-wide aluminium strip (green), which is kept electrically grounded.
Quantum dots on the left and right are created using the finger gates (blue) and the ABS energy is controlled by the central ABS gate (purple).
The biases $V_\mathrm{L}$ and $V_\mathrm{R}$ applied to the left and right leads can be varied independently.
The currents $I_\mathrm{L}$ and $I_\mathrm{R}$ in the left and right leads are measured simultaneously. 
We define a positive current as the flow of electron charge from the leads to the superconductor. 

\begin{figure*}[t!]
\includegraphics[width=1\textwidth,keepaspectratio=true]{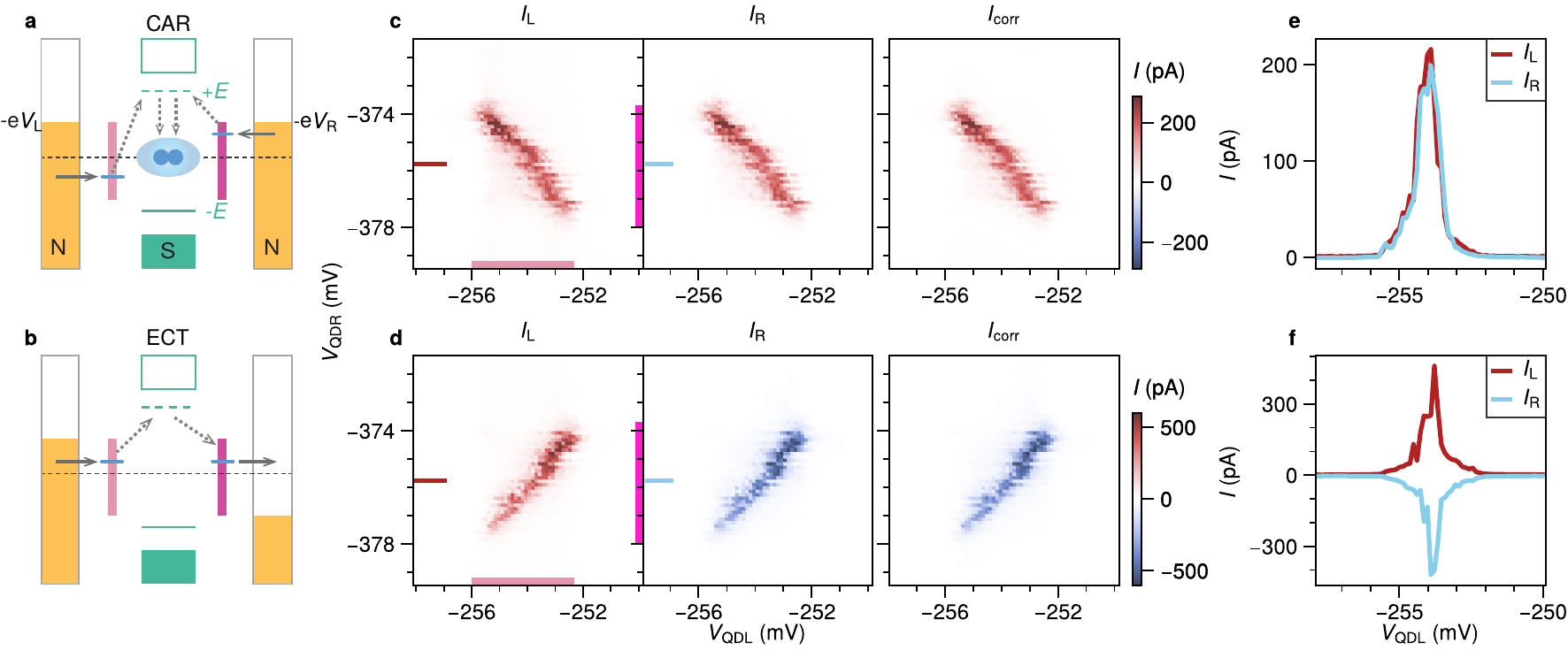}
\caption{
\textbf{Correlated CAR and ECT signals.} Diagrams of the transport cycles for \textbf{(a)} CAR and \textbf{(b)} ECT. Blue lines indicate the energies of the QD levels required for transport via the ABS at energy $\pm E$.
Purple bars mark the energy window in which transport is allowed, corresponding to the marked regions in the measurement panels (\textbf{c} and \textbf{d}).
\textbf{(c)} Charge stability measurement of QDL and QDR with $V_\mathrm{L}$ = $V_\mathrm{R}$ = $\SI{-120}{\upmu V}$ taken at $V_\mathrm{ABS}$ = \SI{-245}{mV}.
Equal currents with the same sign are observed at the left ($I_\mathrm{L}$) and right ($I_\mathrm{R}$) leads only when the QD energy levels are anti-aligned, as expected for CAR.
\textbf{(d)} Repeated measurement, but with $V_\mathrm{L}$ = $-V_\mathrm{R}$ = $\SI{-120}{\upmu V}$. 
Equal currents with opposite sign are observed only when the QDs are aligned in energy, as expected for ECT. 
The correlated currents $I_\mathrm{corr}$ are calculated from $I_\mathrm{L}$ and $I_\mathrm{R}$ as described in the main text. Exemplary line traces at $V_\mathrm{QDR}$ = \SI{-375}{mV} for CAR and ECT are plotted in \textbf{(e)} and \textbf{(f)} respectively.
}
\label{fig2}
\end{figure*}

First, the two innermost finger gates are used to define tunneling barriers on either side of the hybrid region. 
\Cref{fig1}c,d show the measured local conductance $G_\mathrm{RR}$ = $\frac{dI_\mathrm{R}}{dV_\mathrm{R}}$ and non-local conductance $G_\mathrm{LR}$ = $\frac{dI_\mathrm{L}}{dV_\mathrm{R}}$ as a function of the ABS gate voltage $V_\mathrm{ABS}$.
The induced gap in the hybrid section is found to be $\Delta_\mathrm{ind}\approx$
$\SI{220}{\upmu eV}$.
The correspondence between $G_\mathrm{RR}$ and $G_\mathrm{LR}$ shows the presence of an extended discrete ABS in the proximitized section.
The observed sign-switching in the non-local signal is typical for an extended ABS probed in a three-terminal measurement~\cite{Gramich2017,Danon_PRL_2020,Poschl2022}.
Next, two quantum dots are created on either side of the proximitized section.
Their electro-chemical potentials are controlled by applied voltages $V_\mathrm{QDL}$ and $V_\mathrm{QDR}$.
The charge stability diagrams of both QDs (\Cref{fig1}e,f) show Coulomb diamonds \textcolor{black}{with clear even-odd spacing}. 
\textcolor{black}{The pair of Coulomb peaks show linear splitting as a function of magnetic field, indicative of a spin-degenerate single orbital level (Fig.~S1)}.
The superconducting gap $\Delta_\mathrm{ind}$ is clearly visible at the charge degeneracy points, indicative of a weak coupling to the \textcolor{black}{proximitized region}~\cite{Deacon_2010_PRB_Kondo,Gramich_2015_PRL}. 
Charging energies of QDL and QDR are \SI{1.9}{meV} and \SI{1.4}{meV} respectively, much larger than the induced superconducting gap.
\vfill
\null

\begin{figure*}[t!]%
\includegraphics[keepaspectratio=true,width = \textwidth]{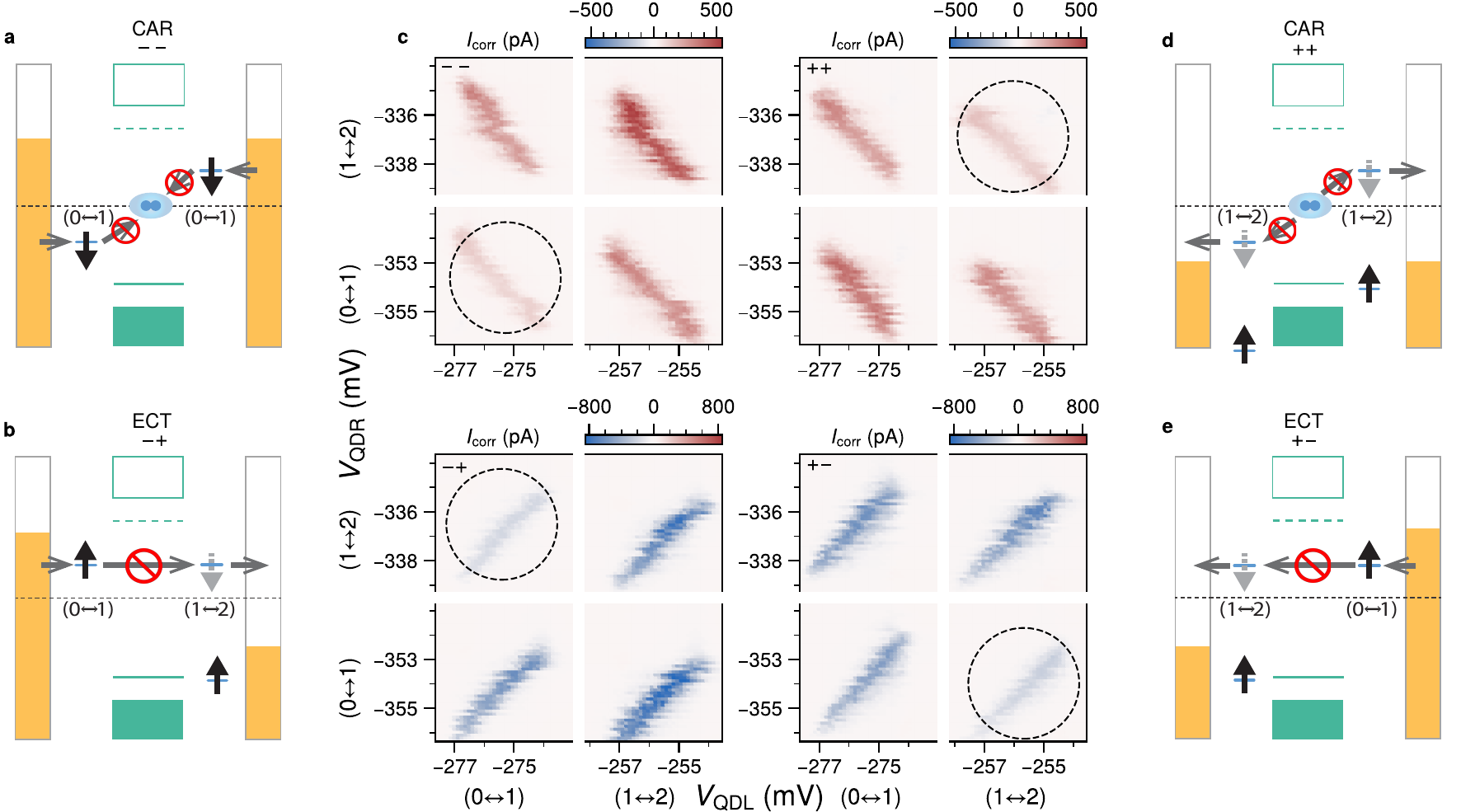}
\caption{\textbf{Spin blockade at zero magnetic field.} Charge stability diagrams are obtained for all four bias polarity combinations, to measure either CAR ($V_\mathrm{L} = V_\mathrm{R}$) or ECT ($V_\mathrm{L} = -V_\mathrm{R}$). 
\textbf{(a),(b),(d)} and \textbf{(e)} show energy diagrams illustrating situations expected to lead to transport blockades. 
Arrows within the dots either represent an already occupied spin state (black), or a state available to be occupied by an incoming electron (grey).
\textbf{(c)} The corresponding measurements of $I_\mathrm{corr}$ plotted against $V_\mathrm{QDL}$ and $V_\mathrm{QDR}$, with applied biases $\lvert V_\mathrm{L}\rvert =$ $\lvert V_\mathrm{R}\rvert=$ $\SI{120}{\upmu V}$ and $V_\mathrm{ABS}$ = \SI{-220}{mV}.
The used bias polarity for each set of measurements is noted in the top left corner.
For each bias configuration a specific transition is suppressed (dashed circles) as a consequence of the blockade depicted alongside the measurement. Gate voltage ranges are interrupted to zoom-in on the relevant ECT and CAR features.}
\label{fig3}
\end{figure*}
\subsection{CAR and ECT}
For CAR, an electron from each of the two leads is simultaneously transferred to the superconductor via an extended ABS to form a Cooper pair (\Cref{fig2}a).
This should therefore result in positively correlated currents in the leads ($I_\mathrm{L} = I_\mathrm{R}$). 
For ECT (\Cref{fig2}b), an electron from the left or right lead tunnels to the opposite lead via the hybrid section, which should thus give rise to negatively correlated currents ($I_\mathrm{L} = -I_\mathrm{R}$). 
As we will show below, by controlling the QD levels and voltage biases, it is possible to distinguish currents arising from ECT and CAR.
Such measurements are shown in \Cref{fig2}c,d.
Here $V_\mathrm{QDL}$ and $V_\mathrm{QDR}$ are each tuned close to a selected charge degeneracy point and the currents $I_\mathrm{L}$ and $I_\mathrm{R}$ are simultaneously measured.
The large charging energies of the dots ensure that each lead strongly prefers accepting or donating a single electron. 
We further ensure that the applied biases are lower in energy than any sub-gap states in the hybridized region, such that local transport is suppressed.
To demonstrate CAR, we set $V_\mathrm{L} = V_\mathrm{R}$ = $\SI{-120}{\upmu V}$ and sweep $V_\mathrm{QDL}$ and $V_\mathrm{QDR}$. 
A finite current is observed only along a line with negative slope, for both $I_\mathrm{L}$ and $I_\mathrm{R}$ (\Cref{fig2}c).
Furthermore, the currents are equal both in magnitude and sign (\Cref{fig2}e).
Converting the gate voltages to electro-chemical potentials ($\mu_\mathrm{L}, \mu_\mathrm{R}$), we confirm that CAR mediated transport occurs when $\mu_\mathrm{L} = -\mu_\mathrm{R}$ (Fig.~S4c).
This is consistent with the requirement that the energies of the electrons forming the Cooper pair must be equal and opposite. 
To demonstrate ECT, we apply biases with opposite polarity ($V_\mathrm{L}$ = $-V_\mathrm{R}$ = $\SI{-120}{\upmu V}$).
Unlike CAR, a finite current is observed only along a line with positive slope (\Cref{fig2}d). This is consistent with energy conservation during ECT, which demands that $\mu_\mathrm{L} = \mu_\mathrm{R}$. 
Furthermore, the currents are now equal in magnitude, but opposite in sign (\Cref{fig2}f). 
Note that when biasing only $V_\mathrm{L}$ or $V_\mathrm{R}$ and grounding the other lead, both ECT and CAR become visible in the charge stability diagram (Fig.~S2).

Importantly, for both CAR and ECT we observe no notable current when the bias and energy conditions are not met, indicating that unwanted local processes are strongly suppressed. 
In combination with strongly correlated currents, this suggests a relatively large signal-to-noise ratio of the CPS process.
To characterize this, we calculate the CPS efficiency and visibility (Fig.~S4). 
Following\cite{Schindele_PRL_2012,Wang2022}, we obtain a combined CPS efficiency above 90~\%, on par with the highest previously reported values~\cite{Schindele_PRL_2012, Wang2022}.
Applying a larger bias that exceeds the sub-gap state energy (but is still below $\Delta_\mathrm{ind}$) results in additional local, non-correlated signals which only depend on a single QD (Fig.~S3) and significantly reduce the CPS efficiency. 

To systematically characterize the CAR and ECT measurements, we calculate the correlated current $I_\mathrm{corr} \equiv \text{sgn}(I_\mathrm{L} I_\mathrm{R})\cdot\sqrt{|I_\mathrm{L}| |I_\mathrm{R}|}$ (\Cref{fig2}c,d)\textcolor{black}{\cite{Wang2022}}. 
It is non-zero only when $I_\mathrm{L}$ and $I_\mathrm{R}$ are both non-zero and thus highlights features mediated by ECT and CAR.
Furthermore the sign of $I_\mathrm{corr}$ clearly distinguishes CAR (always positive) from ECT (always negative). 

\subsection{Zero field spin blockade}
\begin{figure*}[t!]%
\includegraphics[width=\textwidth]{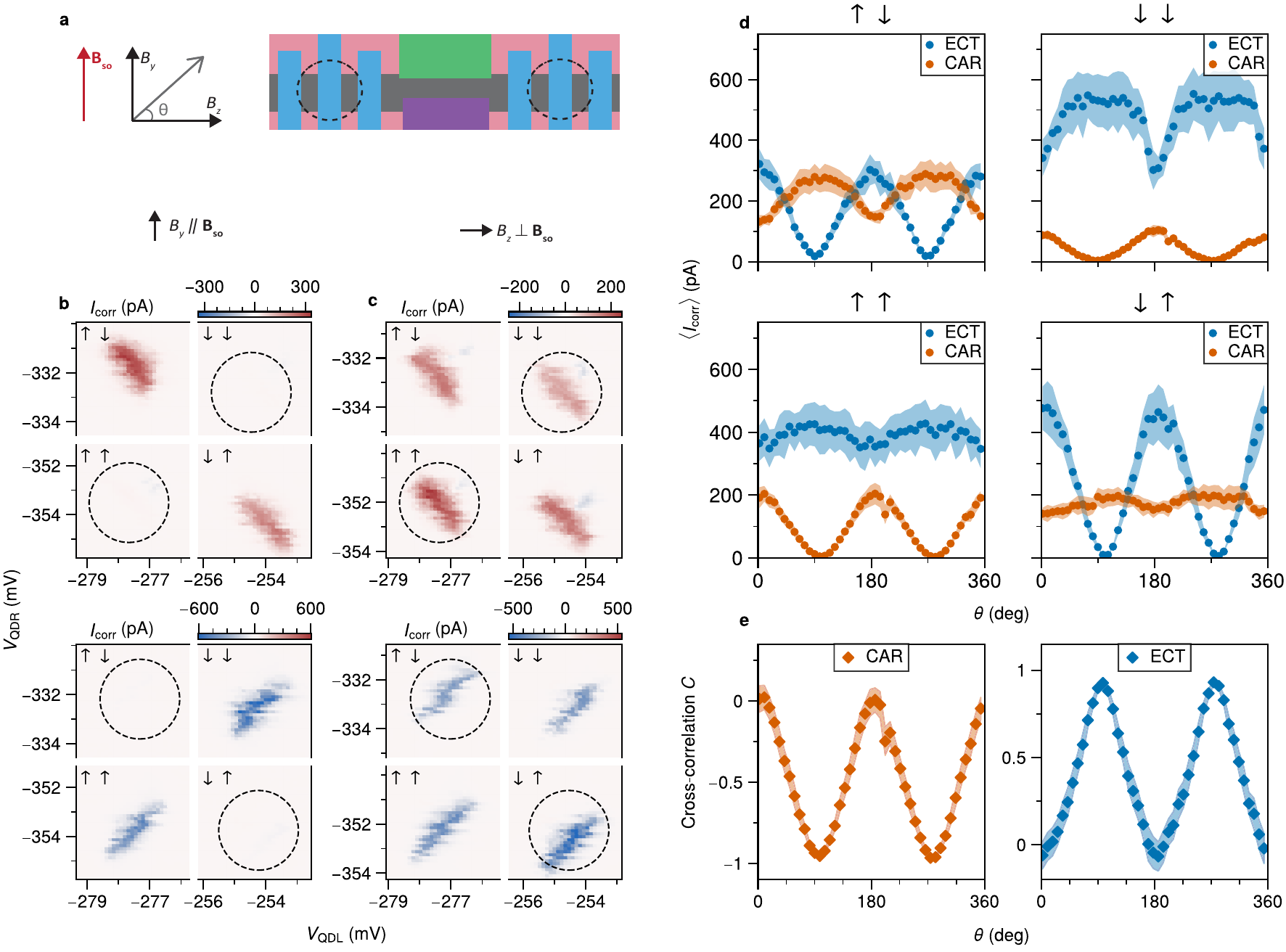}

\caption{\textbf{CAR and ECT at finite magnetic field}. 
\textbf{(a)} A schematic showing the coordinate system of the applied magnetic field with respect to the device.
\textbf{(b)} Measurement of $I_\mathrm{corr}$ for CAR~(top) and ECT~(bottom) with $\mathrm{\mathbf{B}}\parallel B_\mathrm{y}$ = \SI{150}{mT} and $V_\mathrm{ABS}$ = \SI{-220}{mV}. Lower biases ($\lvert V_\mathrm{L}\rvert,\lvert V_\mathrm{R}\rvert~$= $\SI{70}{\upmu V}$) are applied to keep the bias window below any sub-gap states, whose energies are pulled down by the finite magnetic field (Fig.~S1). Equal-spin CAR and opposite-spin ECT are fully suppressed (circled).
\textbf{(c)} Measurement of $I_\mathrm{corr}$ with $\mathrm{\mathbf{B}}\parallel B_\mathrm{z}$ = \SI{150}{mT}. The blockades in \textbf{(b)} have been clearly lifted.
\textbf{(d)} Angle-dependence of $\langle I_\mathrm{corr} \rangle$ for the different spin channels for a full rotation of the magnetic field in the y-z plane.
The ($-$,+) and ($+$,$+$) bias configurations are used for ECT and CAR respectively.
Each data point represents a single charge stability diagram for a specific spin channel. \textcolor{black}{The data extraction procedure is described in Fig.~S7.}
\textbf{(e)} The calculated spin cross-correlation (as defined in the text) of CAR and ECT, derived from (\textbf{d}).}
\label{fig4}
\end{figure*}
In the absence of a magnetic field, the orbital levels of the QDs are spin-degenerate.
Therefore, if the dot has an even number of electrons, the first electron to occupy the next orbital (a transition denoted as $0\leftrightarrow1$) can be either spin-up or spin-down. 
However, to add the second electron ($1\leftrightarrow2$), the Pauli exclusion principle requires it to have an opposite spin. 
The effect of this spin-filling rule leads to a blockade of transport, which depends on the nature of the underlying process.

We first focus on ECT in the ($-$,+) bias configuration, denoting that a negative bias is applied to the left lead and a positive bias is applied to the right lead (\Cref{fig3}b). 
When the QDs are tuned to the ($0\leftrightarrow1$, $1\leftrightarrow2$) transition, a situation can arise where the left QD is occupied with e.g. a spin-up electron (coming from the left lead), whereas the right QD can only accept a spin-down electron (since the spin-up state has already been occupied). 
At this point transport from left to right is blocked, analogous to the well-known Pauli blockade in double quantum dots~\cite{VdWiel_RMP_double_QDs_2022}. 
This spin blockade is clearly seen when the QDs are tuned over successive charge transitions. 
In \Cref{fig3}c we see that the ECT current is suppressed for the ($0\leftrightarrow1$, $1\leftrightarrow2$) transition. 
Reversing the bias polarities to (+,$-$), a similar blockade is observed for the ($1\leftrightarrow2$, $0\leftrightarrow1$) transition, as expected (see \Cref{fig3}e).

In the ($-$,$-$) configuration, only CAR mediated transport can occur and we find a suppression in CAR current for the ($0\leftrightarrow1$, $0\leftrightarrow1$) transition.
This is a direct consequence of the Cooper pairs in an s-wave superconductor having a singlet pairing. 
Thus, for transport to occur, each QD must donate an electron of \textit{opposite} spin in order to create a singlet Cooper pair in the superconductor. 
Transport is therefore blocked when both dots are occupied by electrons with the same spin (\Cref{fig3}a).
Finally, in the (+,+) configuration a blockade is expected for the ($1\leftrightarrow2$, $1\leftrightarrow2$) transition (\Cref{fig3}d), as observed in the measurements. 
Qualitatively similar measurements of spin blockade for CAR and ECT are presented for another device~(Fig.~S8).
We note that a finite amount of current remains for each blockaded transition, indicating the presence of a spin-relaxation mechanism in our system. 
The hyperfine interaction is one such mechanism that can lift the Pauli blockade \cite{npstevan_2010_SOC_blockade,Danon_2009_SOC_blockade_theory}. 
We confirm this by applying a magnetic field to \textcolor{black}{suppress the spin-mixing due to the hyperfine interaction,} and find that 35~mT is sufficient to fully suppress the remaining current (Fig.~S5).

\subsection{Singlet and triplet ECT/CAR}
The spin degeneracy of the QD levels is lifted by applying a magnetic field, allowing us to operate them as spin-filters (Fig.~S1).
When the Zeeman splitting exceeds $\lvert eV_\mathrm{L}\rvert,\lvert eV_\mathrm{R}\rvert$, only spin-up ($\uparrow$) electrons are involved in transport at a ($0\leftrightarrow1$) transition and only spin-down ($\downarrow$) at a ($1\leftrightarrow2$) transition.
In the absence of spin-orbit coupling, CAR is only expected to occur when both QDs are tuned to host electrons with opposite spin. 
The opposite applies to ECT, where a current is only expected when the QDs are tuned to receive electrons with equal spin.

As shown in \Cref{fig4}b, when an in-plane field of 150~mT is applied along $B_\mathrm{y}$ (i.e. perpendicular to the channel), CAR current is only present in the quadrants where the electrons have opposite spins ($\uparrow \downarrow$ and $\downarrow \uparrow$) and completely suppressed for the equal-spin ($\uparrow \uparrow$ and $\downarrow \downarrow$) configuration.  
Similarly, no current is detected for opposite-spin ECT, while transport is allowed for equal-spin ECT. 
This spin-dependent transport indicates that the direction of the spin-orbit field $\mathrm{\mathbf{B}}_\mathrm{SO}$ is along $B_\mathrm{y}$, making spin a good quantum number. 
This is also consistent with the expected Rashba spin-orbit interaction in a quasi-1D channel with momentum along the z-direction and electric field perpendicular to the 2DEG plane.
Applying the magnetic field perpendicular to $\mathrm{\mathbf{B}}_\mathrm{SO}$ (i.e. along $B_\mathrm{z}$), a spin-up electron may acquire a finite spin-down component, due to the spin-orbit interaction.
The consequence of this can be seen in \Cref{fig4}c, where we now observe sizeable currents for equal-spin CAR and opposite-spin ECT.
The full evolution of the spin-specific ECT and CAR currents can be obtained by performing an in-plane rotation of the magnetic field (\Cref{fig4}d). 
The averaged amplitudes of equal-spin CAR and opposite-spin ECT currents $\langle I_\mathrm{corr} \rangle$ are found to oscillate smoothly between full suppression at $\theta \approx \SI{90}{\degree}$ and $\SI{270}{\degree}$ ($\mathrm{\mathbf{B}} \parallel \mathrm{\mathbf{B}}_\mathrm{SO}$), and their maximum strength at $\theta \approx \SI{0}{\degree}$ and $\SI{180}{\degree}$ ($\mathrm{\mathbf{B}}\perp \mathrm{\mathbf{B}}_\mathrm{SO}$).
This result does not depend on a specific choice of orbitals in the QDs (Fig.~S6). 

The ability to accurately resolve the spin of the electrons in CPS is particularly relevant in the context of entanglement witnessing. An important metric capturing this, is the spin cross-correlation~\cite{Braunecker2013,Klobus2014}. As described in \cite{Bordoloi2022}, we calculate the spin cross-correlation from the measured currents as:
\begin{equation}
    C=\frac{(I^{\uparrow\uparrow}+I^{\downarrow\downarrow}-I^{\uparrow\downarrow}-I^{\downarrow\uparrow})}{({I^{\uparrow\uparrow}+I^{\downarrow\downarrow}+I^{\uparrow\downarrow}+I^{\downarrow\uparrow}})}
\end{equation} 
and plot it for both CAR and ECT as a function of $\theta$ (\Cref{fig4}e). \textcolor{black}{$I^{ij}$ corresponds to the average correlated current  $\langle I_\mathrm{corr} \rangle$ associated with each spin configuration, where $i,j\in \{\uparrow,\downarrow\}$.}
$C = \pm1$ when there is a perfect correlation or anti-correlation between the spins of electrons entering the QDs. In contrast, $C=0$ when the probabilities of equal-spin and opposite-spin transport become equal.
When $\mathrm{\mathbf{B}} \parallel \mathrm{\mathbf{B}}_\mathrm{SO}$ we obtain a value of $C$  = $-0.96$ for CAR, demonstrating a nearly perfect singlet pairing between the QDs. Similarly, for ECT $C$  = $+0.93$ is obtained.
When $\mathrm{\mathbf{B}} \perp \mathrm{\mathbf{B}}_\mathrm{SO}$, $C$ reaches close to 0 for both CAR and ECT, stressing that the triplet \textcolor{black}{component} can be tuned to be of similar magnitude to the conventional singlet pairing.

In conclusion, we have used quantum dot-superconductor hybrids to demonstrate highly efficient Cooper pair splitting in a two-dimensional semiconductor platform.
Using spin-polarized quantum dots, we performed spin-selective measurements of ECT and CAR and showed that the strong spin-orbit interaction in ternary 2DEGs results in comparable strengths of singlet and triplet \textcolor{black}{correlations between the quantum dots}.
Finally, through magnetic field rotations, we showed that it is possible to obtain equal amplitudes of ECT and CAR, establishing 2DEGs as an ideal platform to study Majorana bound states in artificial Kitaev chains.

\section{Discussion}
The demonstration of \textcolor{black}{singlet and triplet correlations with Cooper pair splitters} in 2DEGs paves the way for more advanced experiments to study entanglement and topological superconductivity. An interesting open question relates to the underlying mechanism that allows for strong triplet CAR in these devices. 
One possibility is for two equal-spin electrons to form a normal s-wave Cooper pair, due to spin precession in the tunnel barriers. 
Another path is that an induced p-wave superconducting pairing arises in the hybrid section, such that two equal-spin electrons form a Cooper pair. 
In order to distinguish these possibilities, we propose to create quasi-1D channels that are bent (rather than straight), resulting in different spin-orbit directions in each arm of the Cooper pair splitter\cite{Braunecker2013,Hels_PRL_2016}.
Such devices are easily implemented in 2DEGs where any arbitrary shape of the channel can be realized simply by altering the design of the depletion gates. 
\textcolor{black}{Given the high fidelity spin correlation we have demonstrated here,} such devices could also be used to detect entanglement by performing a Bell test with electrons from a Cooper pair~\cite{Braunecker2013}. 

Finally, the recent realization of a minimal Kitaev chain~\cite{Dvir2022} opens up several possibilities to systematically study Majorana bound states (MBSs). In this regard the 2DEG platform is again particularly suitable.
\textcolor{black}{It readily allows for extending these measurements to multi-site QD chains, whereby the flexibility of the 2DEG would allow for the simultaneous measurement of density of states at the edges and in the bulk.}
Furthermore, one could use these chains to perform tests of non-Abelian exchange statistics via braiding experiments~\cite{Alicea2011,Aason_milestone_2016}, which necessarily require a 2D platform.

\clearpage

\section{Methods}
\subsection{Fabrication}
Device 1 (main text) and Device 2 (supplementary) were fabricated using techniques described in detail in \cite{CMohle_Controlled_ABSs_2022}.
A narrow aluminum strip is defined in an InSbAs-Al chip by wet etching, followed by the deposition of two normal Ti/Pd contacts. 
After deposition of \SI{20}{nm} AlOx via atomic layer deposition (ALD), the two depletion gates are evaporated. Following a second ALD (\SI{20}{nm} AlOx) Ti/Au gates are evaporated in order to define the QDs and tune the ABS energy. 

\subsection{Measurements}
All measurements are performed in a dilution refrigerator with a base temperature of \SI{20}{mK}. Magnetic fields are applied using a 3D vector magnet. The alignment of the magnetic field with respect to the device is expected to be accurate within $\pm\SI{5}{\degree}$. Transport measurements are performed in DC using a three-terminal set-up, where the aluminum is electrically grounded ~(\Cref{fig1}b). 
Current amplifier offsets are determined by the average measured current when both dots are in Coulomb blockade.
CAR and ECT processes can be observed over a wide range of $V_\mathrm{ABS}$ voltages. 
Once a $V_\mathrm{ABS}$ setting was found with both strong CAR and ECT currents, it was kept at a constant value throughout the rest of the measurements.
Further care was taken to implement the same orbitals in both QDs for all presented measurements in the main text. 
The mismatch between exact $V_\mathrm{QDR}$ and $V_\mathrm{QDL}$ values at which ECT and CAR are observed is due to gate instabilities, causing a drift of charge degeneracy points over a period of time.
Therefore, the field rotation measurement in \Cref{fig4}e was performed multiple times.
No quantitative difference was observed between measurements.
Presented data was selected due to high stability of the QDs over the course of the measurements. 

Overall, we have measured four fully functional devices at the time of writing this manuscript, all of which have produced highly efficient CAR and ECT mediated by extended ABSs. For three of these devices we have performed magnetic field rotations and observed angle-dependent oscillations of ECT and CAR currents.

\section{Data availability}
Raw data and analysis scripts for all presented figures are available at \url{https://doi.org/10.5281/zenodo.7311374}.

%


\section{Acknowledgments}
We thank T.~Dvir, G.~Wang,  C.~-X.~Liu, M.~Wimmer, C.~Prosko, P.~Makk, R.~Aguado, D. Xu, and L.~P.~Kouwenhoven for valuable discussions and for providing comments on the manuscript. The research at Delft was supported by the Dutch National Science Foundation (NWO) and a TKI grant
of the Dutch Topsectoren Program. The work at Purdue was funded by Microsoft Quantum. 

\section{Author contributions}

Q.~W. and S.~L.~D.~t.~H. fabricated and measured the devices. I.~K. contributed to the device design and optimization of fabrication flow. MBE growth of the semiconductor heterostructures and the characterization
of the materials was performed by D.~X., and C.~T. under the supervision of M.~J.~M. The manuscript was written by Q.~W., S.~L.~D.~t.~H., and S.~G., with inputs from all coauthors. S.~G. supervised the experimental work in Delft.

\section{Competing interests}
\textcolor{black}{The authors declare no competing interests.
}

\end{document}


\title{Triplet correlations in Cooper pair splitters realized in a two-dimensional electron gas}
\author{Qingzhen Wang}
\altaffiliation{These authors contributed equally to this work.}
\affiliation{QuTech and Kavli Institute of Nanoscience, Delft University of Technology, Delft, 2600 GA, The Netherlands}
\author{Sebastiaan L.D. ten Haaf}
\altaffiliation{These authors contributed equally to this work.}
\affiliation{QuTech and Kavli Institute of Nanoscience, Delft University of Technology, Delft, 2600 GA, The Netherlands}

\author{Ivan Kulesh}
\altaffiliation{These authors contributed equally to this work.}
\affiliation{QuTech and Kavli Institute of Nanoscience, Delft University of Technology, Delft, 2600 GA, The Netherlands}

\author{Di~Xiao}
\affiliation{Department of Physics and Astronomy, Purdue University, West Lafayette, 47907, Indiana, USA}

\author{Candice Thomas}
\affiliation{Department of Physics and Astronomy, Purdue University, West Lafayette, 47907, Indiana, USA}

\author{Michael J. Manfra}
\affiliation{Department of Physics and Astronomy, Purdue University, West Lafayette, 47907, Indiana, USA}
\affiliation{Elmore School of Electrical and Computer Engineering, ~Purdue University, West Lafayette, 47907, Indiana, USA}
\affiliation{School of Materials Engineering, Purdue University, West Lafayette, 47907, Indiana, USA}
\affiliation{Microsoft Quantum Lab, West Lafayette, 47907, Indiana, USA}

\author{Srijit Goswami}\email{s.goswami@tudelft.nl}
\affiliation{QuTech and Kavli Institute of Nanoscience, Delft University of Technology, Delft, 2600 GA, The Netherlands}

\maketitle

\begin{figure*}
    \centering
       \includegraphics[width = 01\textwidth]{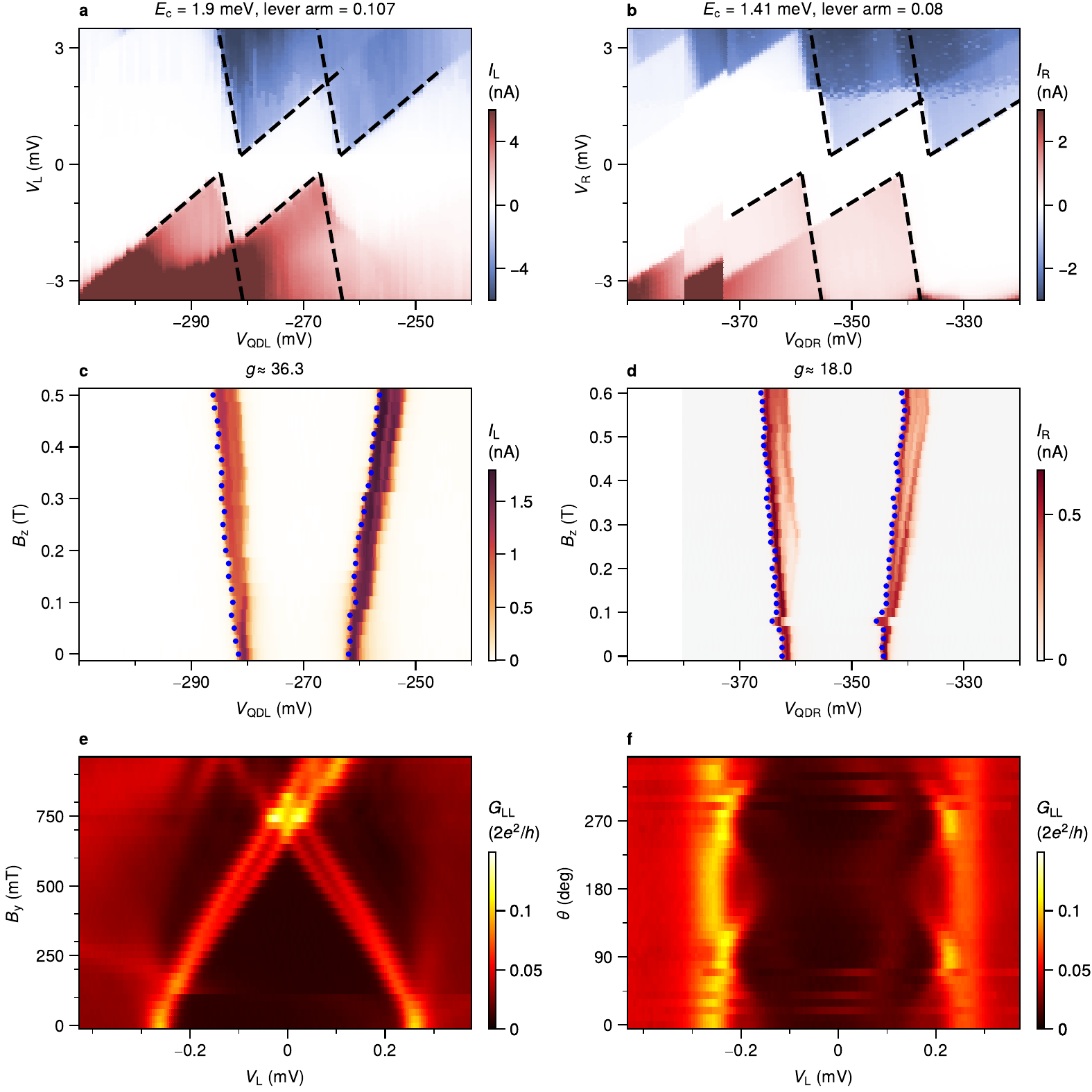}
        \caption{\textbf{Quantum dot and ABS characterization.} Extraction of charging energy $E_\mathrm{c}$ and lever arm for (\textbf{a}) QDL and (\textbf{b}) QDR from Coulomb diamonds shown in Fig~1 of the main text. Evolution of Coulomb peaks with $B_\mathrm{z}$ for (\textbf{c}) QDL and (\textbf{d}) QDR shows Zeeman splitting causing the separation between the resonances to increase. Bias voltages are $(V_\mathrm{L},V_\mathrm{R})$ = (\SI{-450}{\upmu V},0) for the left measurement, and (0, \SI{-350}{\upmu V}) for the right measurement. $g$-factors of -36 and -18 are derived for the QDL and QDR respectively. 
        \textbf{(e)} Measured local conductance $G_\mathrm{LL}$ of the hybrid section with increasing magnetic field $B\parallel B_y$ at $V_\mathrm{ABS}$ = \SI{-245}{mV}. \textbf{(f)} Field rotation in the y-z plane with a fiedld magnitude of \SI{100}{mT}, showing a slight anisotropy of the ABS energy.}
        \label{suppl:dot_zeeman}
    
\end{figure*}

\begin{figure*}[h!]
       \includegraphics[width = \textwidth ]{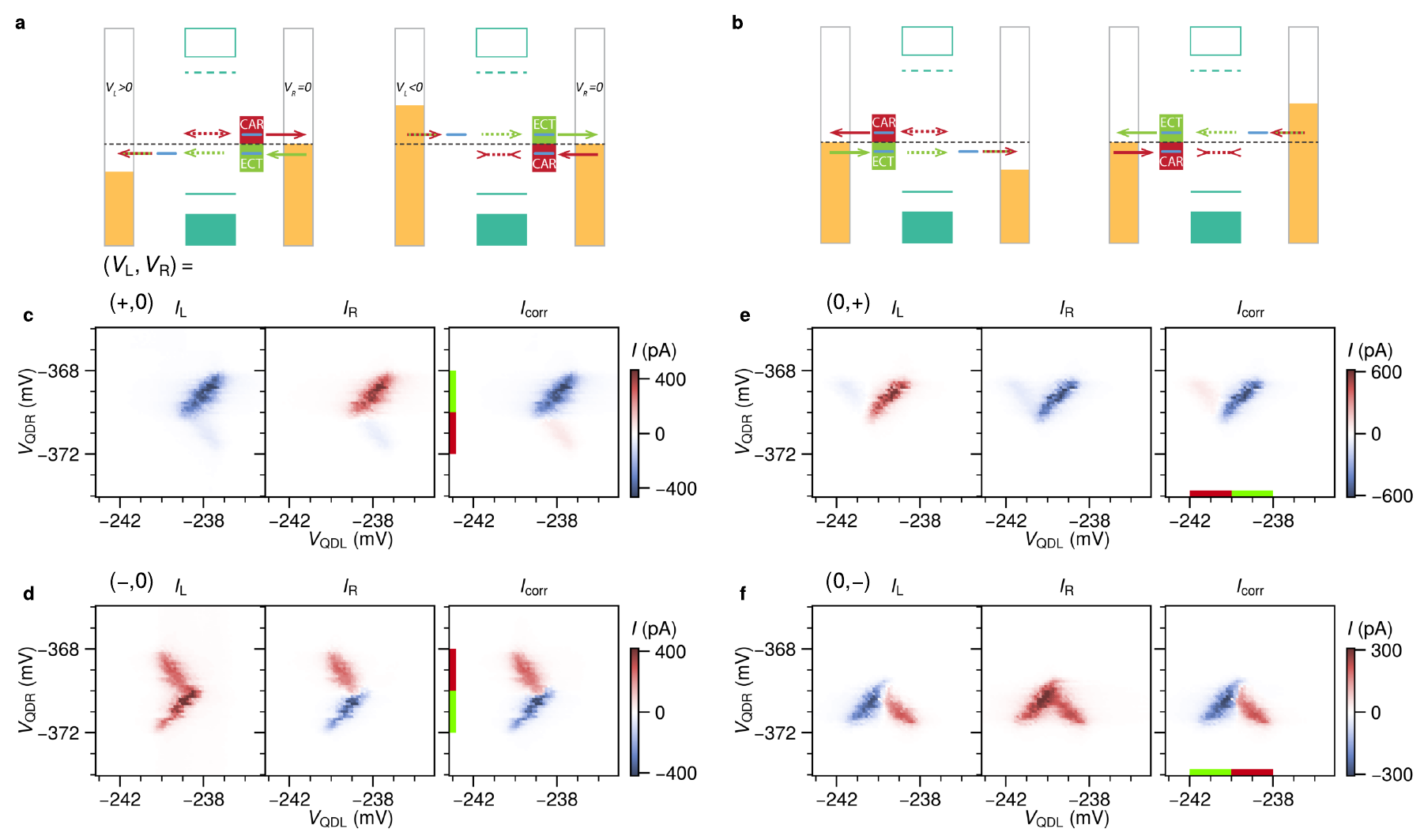}
        \caption{\textbf{Simultaneous measurement of ECT and CAR}. Energy diagrams for the expected transport when biasing only the \textbf{(a)} left  or \textbf{(b)} right side of the device. \textbf{(c)}-\textbf{(f)} Corresponding measurements with the bias polarities labelled in the top left corners. Two distinct features with opposite slope appear in each charge stability diagram. This is because both ECT and CAR can now occur, depending on the position of the QD levels with respect to the grounded side. Red (CAR) and green (ECT) bars indicate the corresponding gate voltage ranges in which each process is allowed. }
        \label{suppl:ECT-CAR-simul}

\end{figure*}

\begin{figure*}[h!]
    \centering
       \includegraphics[width = \textwidth ]{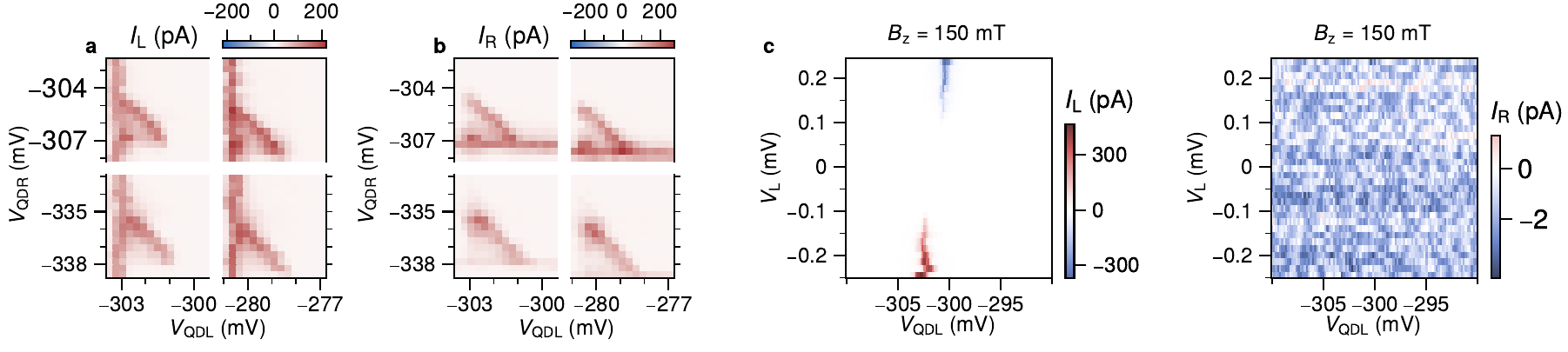}
        \caption{\textbf{Filtering local transport processes.} A key requirement for isolating CAR and ECT is the ability to filter out local processes. For the measurements presented in the main text it was ensured that the applied biases remained below the energy of any subgap states. With applied biases of $-$120~$\upmu$V on both sides and an external magnetic field $B_z$  = 150 m$T$, additional currents appear in \textbf{(a)} $I_\mathrm{L}$ and \textbf{(b)} $I_\mathrm{R}$ on top of the currents arising through CAR. 
        Fixing $V_\mathrm{QDR}$ at $-$320~mV (i.e., putting the right dot off-resonance), $I_\mathrm{L}$ and  $I_\mathrm{R}$ are measured as a function of the left bias $V_\mathrm{L}$ and left plunger gate $V_\mathrm{QDL}$. (\textbf{c}) Current arising through local processes appears once the applied bias exceeds the ABS energy. No current is detected in  $I_\mathrm{R}$.}
        \label{suppl:localAndreev}
    
\end{figure*}

\begin{figure*}[h!]
\centering
\includegraphics[width = \textwidth ]{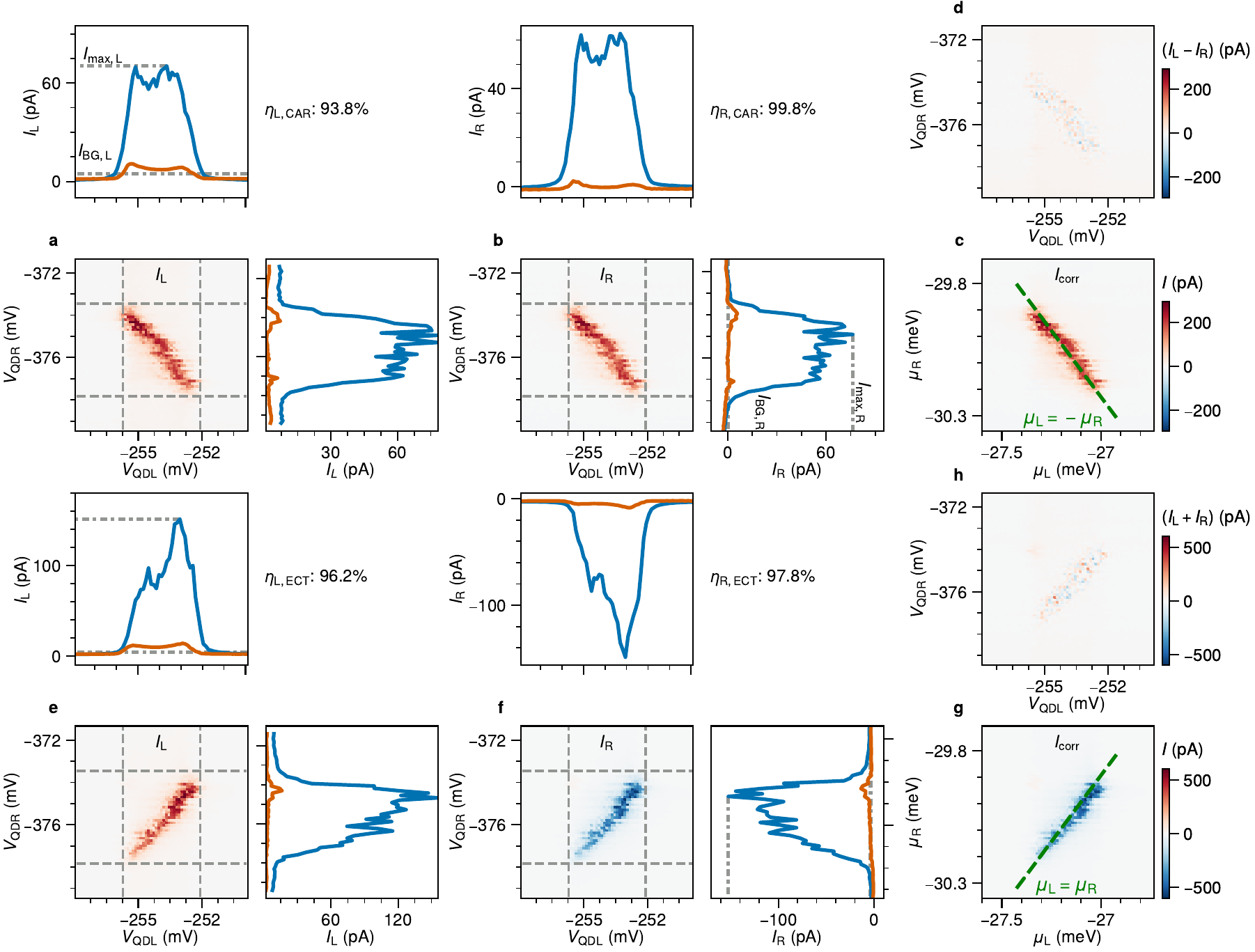}
\caption{\textbf{Cooper pair splitting efficiency.} Analysis of the data presented in Fig.~2. \textbf{(a,b)} Measured currents $I_\mathrm{L}$ and $I_\mathrm{R}$ arising from CAR. The top panel shows $I_\mathrm{L}$ averaged over the $V_\mathrm{QDR}$ range indicated by the horizontal grey lines (blue) and the background trace averaged outside the horizontal grey lines (orange). The right panel shows $I_\mathrm{L}$ averaged over the $V_\mathrm{QDL}$ range indicated by the vertical grey lines and the background trace averaged outside the vertical grey lines. The efficiency of each junction is defined as $\eta_j$ = $1 - \frac{I_{BG,j}}{ I_{max,j}}$ where the $I_{BG,j}$ is the average value of the background trace and $I_{max,j}$ is the maximum of the averaged $I_j$.
We extract an $\eta$ of about 94\% for the left junction and 99\% for the right junction. Taking the product this gives a combined efficiency of $\eta_L \eta_R$ of about 93\%.
Using the lever arms extracted from the Coulomb diamonds (Fig.~S1), the correlated current $I_\mathrm{corr}$ is plotted as a function of $\mu_\mathrm{L}$ and $\mu_\mathrm{R}$ in \textbf{(c)}. The dashed line corresponds to $\mu_\mathrm{L} = -\mu_\mathrm{R}$, confirming that transport occurs when the dot levels are anti-aligned. Calculating $I_\mathrm{L} - I_\mathrm{R}$ shows very small remaining current, verifying $I_\mathrm{L}$ =  $I_\mathrm{R}$ \textbf{(d)}.
\textbf{(e,f)} Similar analysis of $I_\mathrm{L}$, $I_\mathrm{R}$ and averaged currents for ECT. The combined efficiency is again found to be about 93\%. \textbf{(g)} $I_\mathrm{corr}$ in $\mu_\mathrm{L}$- $\mu_\mathrm{R}$ space, together with the line where $\mu_\mathrm{L} = \mu_\mathrm{R}$, showing that transport takes place when the QD levels are aligned. \textbf{(h)} Plotting $I_\mathrm{L}$ + $I_\mathrm{R}$ shows again that little signal remains, highlighting that $I_\mathrm{L} = - I_\mathrm{R}$ for ECT.}
\label{suppl:efficiency}
\end{figure*}

\begin{figure*}[h!]
\centering
\includegraphics[width = \textwidth ]{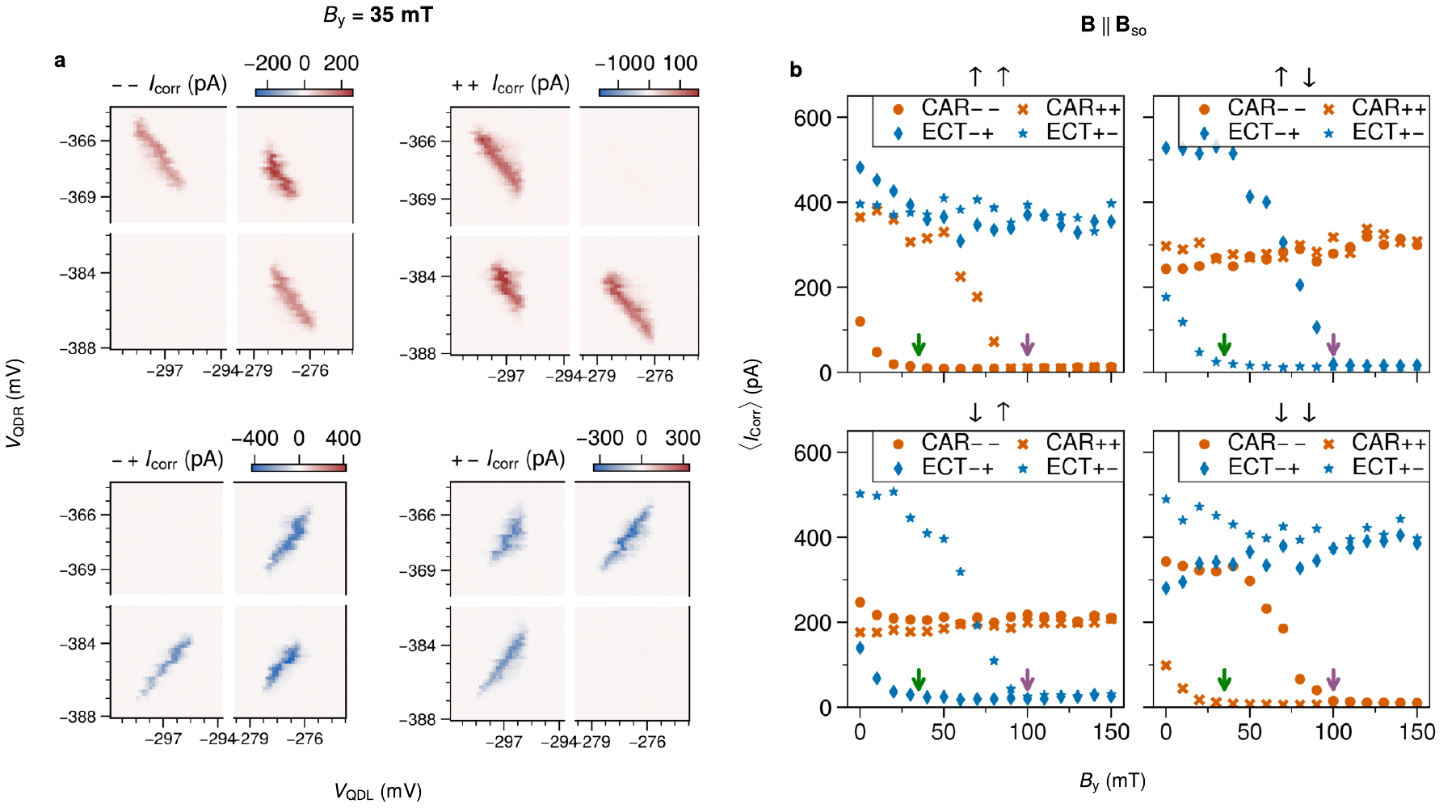}
\caption{\textbf{Field evolution of ECT and CAR.} 
\textbf{(a)} The charge stability diagrams at $B_\mathrm{y}$ = 35~mT, of the same resonances used in the main text, show that the weakened transitions in each bias configuration (circled in Fig.~2c) are now completely blocked. This is interpreted as a result of  the magnetic field overcoming the hyperfine interaction in the QDs.
\textbf{(b)} Spin-filtered measurements of $\langle I_\mathrm{corr}\rangle$ for ECT and CAR as a function of magnetic field $B_\mathrm{y} \parallel \mathbf{B_\mathrm{SO}}$, for each bias configuration. Two transitions are present in each quadrant. Above roughly 35~mT (green arrows) the Pauli-blockaded process in each quadrant becomes fully suppressed. Above 100~mT (purple arrows) the Zeeman splitting exceeds the applied biases ($\lvert$100 $\upmu$V$\rvert$) such that only the spin-preserving processes remain (i.e. opposite-spin CAR and equal-spin ECT).}
    \label{suppl:hyperfine_interaction}
\end{figure*}

\begin{figure*}[h!]
    
\includegraphics[width = \textwidth ]{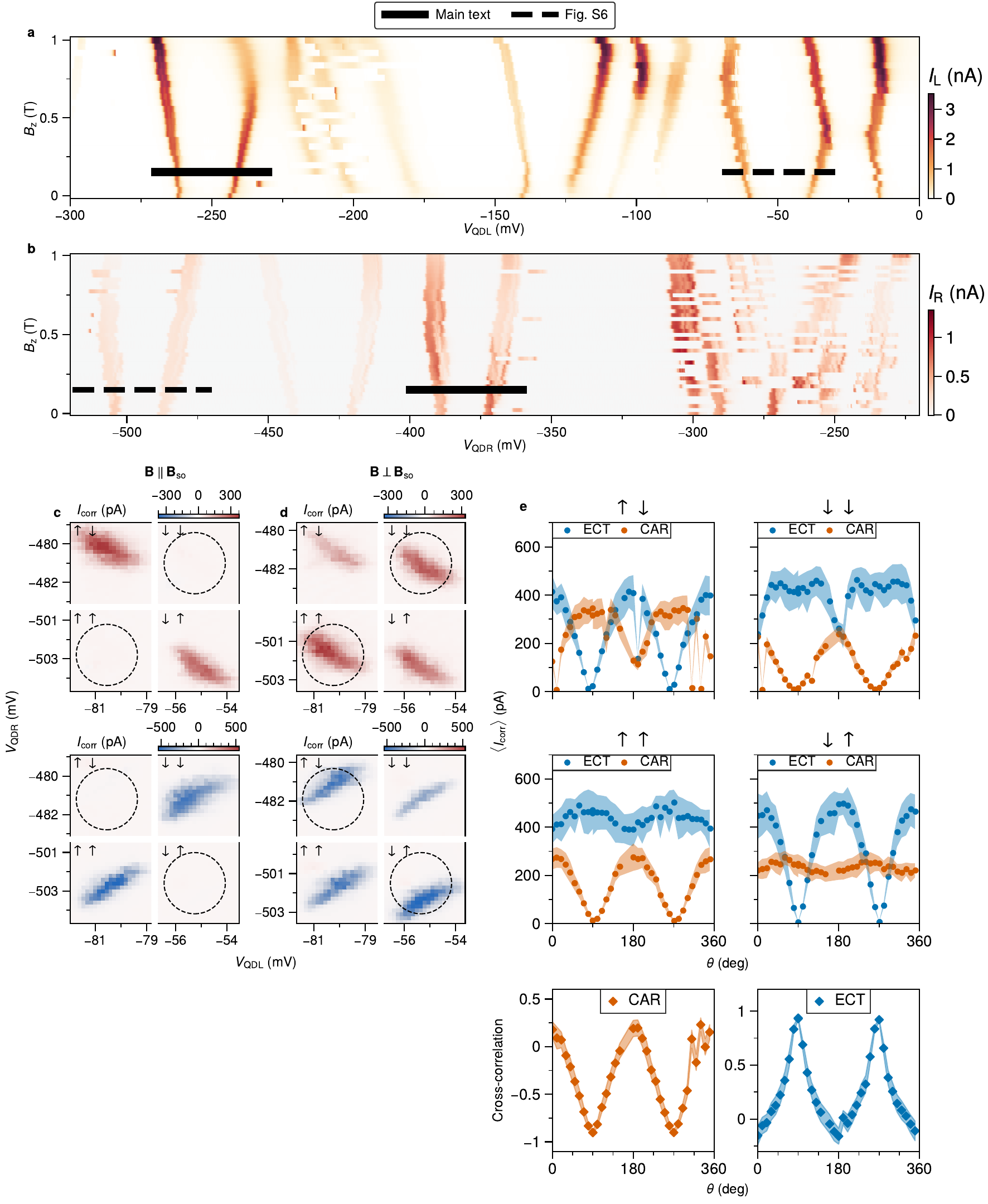}
        \caption{\textbf{Field angle dependence with another pair of QD resonances.} \textbf{(a,b)} Field evolution of Coulomb resonances in the QDs over an extended range of $V_\mathrm{QDL}$ and $V_\mathrm{QDR}$. The effect of Zeeman splitting is observed for multiple orbitals in each QD. Solid black lines indicate the orbitals used in the main text, while the dashed lines indicate different orbitals used for data presented here. \textbf{(c,d)} Repetition of measurements presented in  Fig.~4(c,d) for the second pair of resonances. Unconventional processes (equal-spin CAR and opposite-spin ECT) are fully suppressed at $B \parallel B_\mathrm{SO}$ and recovered at $B \perp B_\mathrm{SO}$. A full field rotation \textbf{(e)} yields similar behaviours of $\langle I_\mathrm{corr} \rangle$ and cross-correlation to the dependence shown in Fig.~4f.}
        \label{suppl:additional_orbital}
\end{figure*}


\begin{figure*}[h!]
\centering
\includegraphics[width = \textwidth]{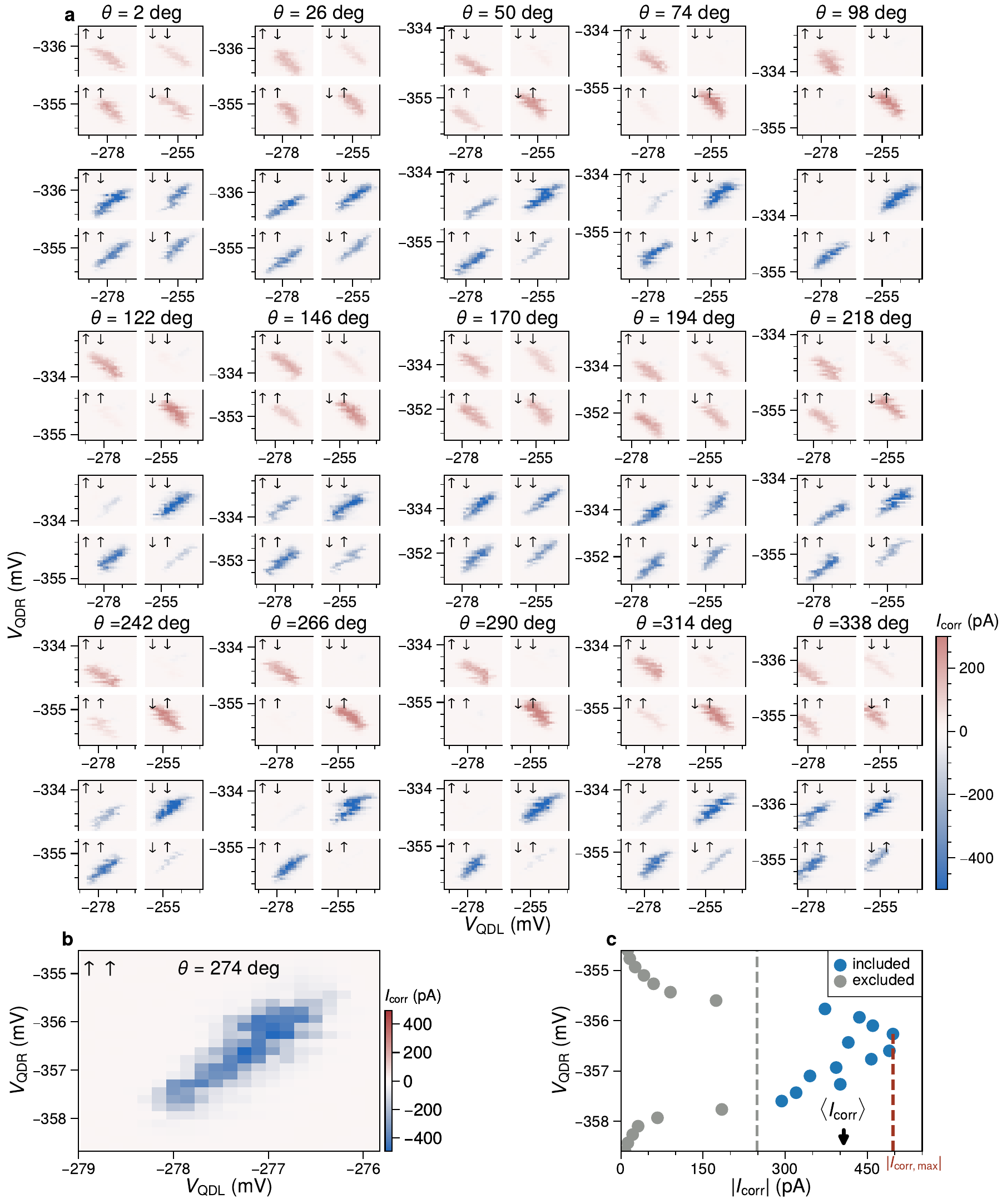}

\caption{\textbf{Raw data from Fig 4f and data extraction.}  \textbf{(a)} Selection of raw data used for extracting the field angle dependence of CAR and ECT rates presented in Fig.~4f. For each angle the four quadrants with different spin configurations of $I_\mathrm{corr}$ are plotted for CAR (top panels) and ECT (bottom panels). A single quadrant is converted to a single data point to quantify the spin-selective rate of CAR and ECT at a specific angle of the magnetic field. The data extraction process is detailed in panel \textbf{(c)} for the measurement of $\uparrow\uparrow$-ECT at $\theta = 274^{\circ} $ (plotted again in \textbf{(b)}). The maximum value of $\lvert I_\mathrm{corr}\rvert$ for each horizontal line-cut is extracted and plotted as a function of $V_\mathrm{QDR}$. To exclude data at the edge of the bias window, a threshold is set at half of the maximally recorded $\lvert I_\mathrm{corr,max}\rvert$. The included values are labeled in blue, from which the average $\langle I_\mathrm{corr}\rangle$ and standard deviation is derived.}
\label{raw_data_Fig4_extraction}
\end{figure*}

\begin{figure*}[h!]
\includegraphics[width = \textwidth ]{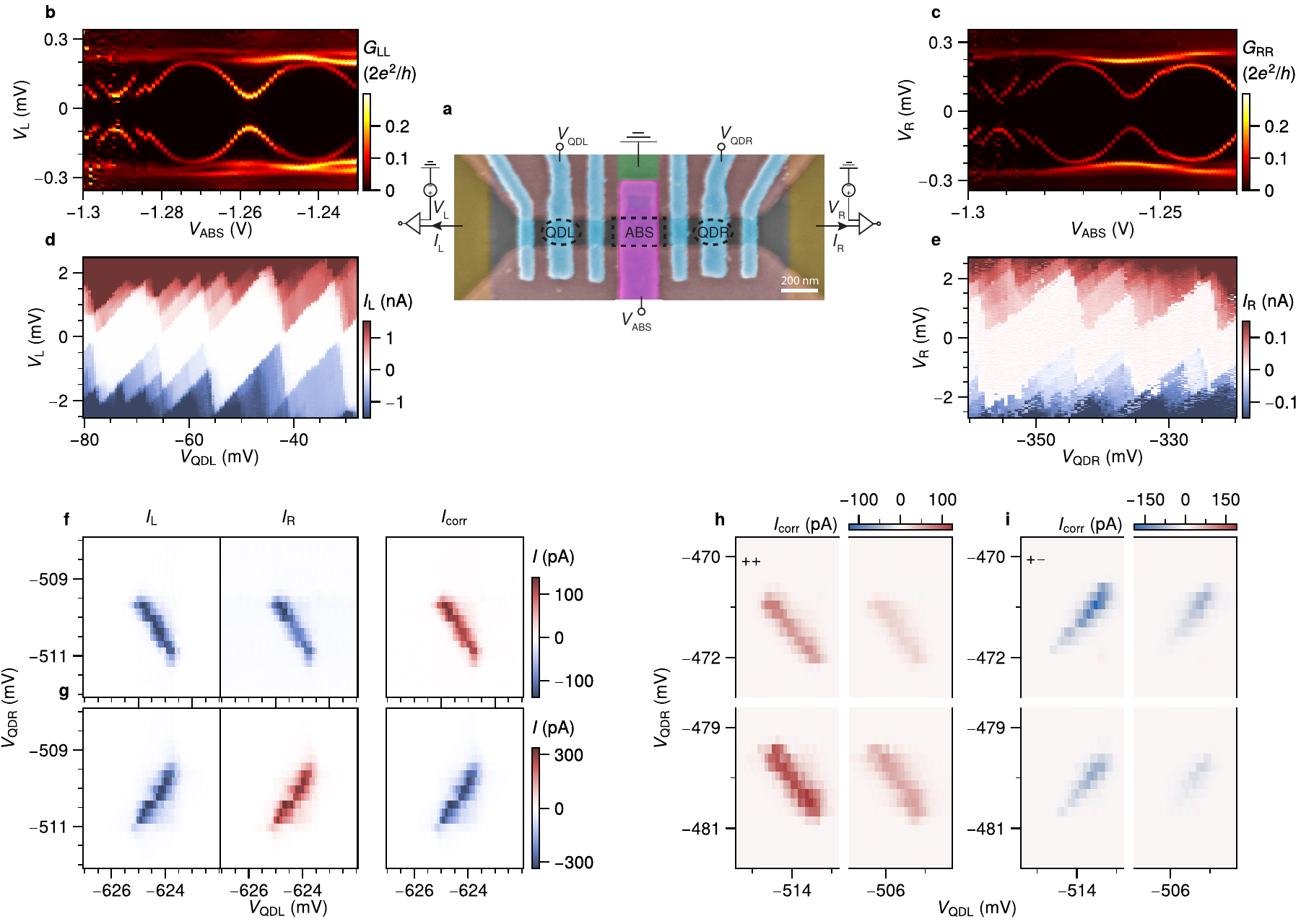}
\caption{\textbf{Measurements on additional device (Device~2).} \textbf{(a)} False-color scanning electron micrograph of Device~2, with the circuit diagram used for three-terminal measurements. Scale bar is \SI{200}{nm}. The distance between the centers of the QDs is about \SI{800}{nm}~(in contrast to \SI{600}{nm} for Device~1 presented in the main text). 
Tunneling spectroscopy measurements of the \textbf{(b)} local conductance $G_\mathrm{LL}$ and \textbf{(c)} $G_\mathrm{RR}$ as a function of $V_\mathrm{ABS}$ show correlated states, indicating the presence of an extended ABSs.
Measured Coulomb diamonds for \textbf{(d)} QDL and \textbf{(e)} QDR. The extracted charging energy is about \SI{1.2}{meV} for both dots. \textbf{(f)},\textbf{(g)} Similar to Fig.~2, measured left current $I_\mathrm{L}$, right current $I_\mathrm{R}$ and the calculated correlated current $I_\mathrm{corr}$, for CAR ($V_\mathrm{L}$  = $V_\mathrm{R}$ = $\SI{100}{\upmu V}$) and ECT ($V_\mathrm{L}$  = $\SI{100}{\upmu V}$ and  $V_\mathrm{R}$ = $\SI{-100}{\upmu V}$ ). Measurements over successive charge transitions reveal the expected spin blockade for \textbf{(h)} CAR (+,+) and \textbf{(i)} ECT (+,$-$) at zero magnetic field.}
\label{suppl:deviceB_part1}
\end{figure*}